\documentclass [11pt,a4paper] {article}

\usepackage[cp1252]{inputenc}

\usepackage{amssymb}
\usepackage{amsmath}
\usepackage{amsfonts,amssymb}

\usepackage[dvips]{graphicx}

\usepackage{bbm}

\DeclareMathAlphabet{\mathpzc}{OT1}{pzc}{m}{it}

\begin{document}

\title{Constructibility of the universal wave function}

\author{Arkady Bolotin\footnote{$Email: arkadyv@bgu.ac.il$} \\ \textit{Ben-Gurion University of the Negev, Beersheba (Israel)}}

\maketitle

\begin{abstract}\noindent This paper focuses on a constructive treatment of the mathematical formalism of quantum theory and a possible role of constructivist philosophy in resolving the foundational problems of quantum mechanics, particularly, the controversy over the meaning of the wave function of the universe. As it is demonstrated in the paper, unless the number of the universe's degrees of freedom is fundamentally upper bounded (owing to some unknown physical laws) or hypercomputation is physically realizable, the universal wave function is a non-constructive entity in the sense of constructive recursive mathematics. This means that even if such a function might exist, basic mathematical operations on it would be undefinable and subsequently the only content one would be able to deduce from this function would be pure symbolical.\\

\end{abstract}

\section{Introduction}\label{Introduction}

\noindent According to Errett Bishop's seminal monograph \textit{Foundations of Constructive Analysis} \cite{Bishop}, the successful formalization of mathematics has fitted space, number, and everything else into a matrix of idealism where even the positive integers have an ambiguous computational existence. Mathematics has evolved into a fine game that has become its own justification.\\

\noindent Interestingly, though, Bishop might have said the same exact words regarding \textit{the evolution of quantum formalism} as well.\\

\noindent Because quantum formalism was originally developed to explain phenomena occurring in the region of single electrons and atoms, within that region quantum formalism has come out \textit{constructive}. This means that every mathematical statement of quantum theory within the given region ultimately expresses the fact that if we perform a finite number of basic mathematical operations in accordance with some procedure, we get certain numerical content, which can afterward be verified experimentally.\\

\noindent Nonetheless, the idea has since germinated to substantially extend the original domain of quantum formalism and apply it to macroscopic objects (such as measurement apparatuses and observers themselves) containing many orders of Avogadro's number of atoms and hence placing far beyond the region where quantum mechanics has been directly tested. But, having applied the mathematical framework of quantum mechanics to the description of macroscopic bodies, quantum theory took a leap from the mathematical entities \textit{constructed in fact} (just as wave functions of a single electron or atom) to \textit{the hypothetically constructible entities} (like macroscopic wave functions).\footnote{Quantum formalism provides no indication as to how to exhibit, or calculate, macroscopic wave functions except the general form of the Schrödinger equation, which, to paraphrase Walter Huckel \cite{Huckel}, has the good view but leads to misunderstanding, because it forces us to think that we have something that we do not have — i.e., a calculated wave function of any given physical system (i.e., of an arbitrary size and complexity).\vspace{5pt}} And upon positing the existence of the universal wave function (i.e., the wave function of the entire universe), quantum theory went even further.\\

\noindent The justification and motivation for extending the original domain of quantum formalism stem from one natural and ``intellectually economical'' assumption: As there is no a priori reason for believing that quantum mechanics may break down when it is applied to systems with a great number of constituent microscopic particles (and hence with a large number of the degrees of freedom), then \textit{there must exist} wave functions that characterize macroscopic bodies and even the entire universe.\\

\noindent On the other hand, since the assumption of the existence of macroscopic wave functions relies on an instance of the law of excluded middle (namely, either macroscopic wave functions can exist or they cannot), the statement of the universal applicability of the quantum formalism cannot be valid within \textit{a constructive existence proof}.\footnote{A constructive existence proof demonstrates the existence of a mathematical object by outlining a method (i.e., an algorithm) of ``constructing'' (that is, computing) this object \cite{Troelstra}.\vspace{5pt}} Specifically, this means that such a statement does not suggest any direct evidence for the existence of the wave function of the universe in the manner that would give us a generic exact algorithm capable of calculating wave functions representing each possible universe (or each possible configuration of geometry and matter field of the universe).\\

\noindent Therefore, despite various concretely constructed approximations of the universal wave function (which, by being the results of drastically simplified toy models, merely refer to allowed values of a single parameter or, at best, a few of them),\footnote{For example, most models in quantum cosmology – an application of quantum theory to the universe as a whole – are based on the approach of the Wheeler–DeWitt equation that combines mathematically the ideas of quantum mechanics and general relativity. This approach is first to restrict the full configuration space of three-geometries called “superspace” to a small number of variables (such as scale factor, inflaton field, etc.) and then to quantize them canonically; consequently, the resultant models are called “minisuperspace models”. More details can be found in \cite{Kiefer04} as well as in the review \cite{Kiefer05}.\vspace{5pt}} the question remains: Can the existence of the wave function, which describes all the degrees of freedom of the universe, be constructively provable, albeit even in principle?\\

\noindent Clearly, if the answer to this question were to be found negative, it would imply that such a function is uncomputable, that is, it does not map computable points (such as integers characterizing microscopic particles of the universe) onto another computable points (such as positive integers that characterize the square absolute value of this function). Along these lines, the negative answer would render the universal wave function an ideal mathematical entity devoid of any empirical meaning.\\

\noindent With the purpose of facilitating the resolution of this important from the ontological point of view question, in the present paper we will elucidate some of the constructivist aspects of the universal wave function.\\

\section{Preliminaries}\label{Preliminaries}

\noindent But first, let us look very briefly over certain features of constructive mathematics that will be referred to bellow in this paper.\\

\noindent Among different varieties of mathematical constructivism (that is, positions with respect to the foundations of mathematics, which demand some form of explicitness of the objects studied, their explicit definability and capability of being viewed as mental constructions – like finitism, predicativism, intuitionism, constructive recursive mathematics (CRM) or Bishop's constructive mathematics (BCM) \cite{Bridges}), those pertaining to the mathematical framework of quantum mechanics (more notably than others) are probably CRM and BCM because they concern with effective calculability of the mathematical entities (e.g., wave functions whose existence is assured or declared by a quantum model). Within these varieties, the notion of ``\textit{existence}'' is taken to be \textit{constructibility}, which is connected with the notion of ``\textit{algorithm}'' to the extent that the adjective ``\textit{constructive}'' becomes equal to the adverb-adjective ``\textit{effectively calculable}'' (or the adjectives ``\textit{computable}'' and ``\textit{recursive}'') \cite{Shanin, Dalen}.\footnote{However, while one can prove that any constructive function is a computable function (for instance, using Kleene's realizability interpretation \cite{Kleene}), it is important to keep the distinction between the notions of constructivism and computability. As explained in \cite{Coquand}, the notion of function in constructive mathematics is a primitive one, which cannot be explained in a satisfactory way in term of recursivity. Even so, unless it brings about a confusion, henceforward in this paper, we will use the adjectives “constructive”, “computable” and “recursive” interchangeably.\vspace{5pt}}\\

\noindent Besides, as regards CRM and BCM, the examination of calculability of the mathematical entities is carried out within the framework of the abstraction of \textit{potential existence (potential realizability)} and under the total exclusion of the idea of \textit{an actual infinity}. Here the abstraction of potential realizability means, e.g., that we may regard addition as \textit{a well-defined operation} (i.e., sensible and definite one) for all natural numbers, since we know how to complete it for arbitrary large numbers \cite{Hilbert}. Therewith, the abstraction of actual infinity involves the acceptance of infinite entities (such as the set of all natural numbers) as given objects \cite{Jane}.\\

\noindent It is worth noticing that providing a method that demonstrates the existence of a mathematical object via calculating this object, a constructive proof does not concern with feasibility (or efficiency) of the method but only with its effectiveness, that is, with its ability to produce a result in a finite number of steps. Consequently, we will not consider computational complexity of constructive procedures in this paper concentrating instead on the existence or non-existence of algorithms for computing wave functions appearing in quantum theory.\\

\section{Non-constructivity of quantum world}\label{Non_constructivity}

\noindent Let us show here that the mathematical foundations of quantum theory are unavoidably non-recursive.\\

\noindent Out of different mathematical formalisms employed in quantum mechanics one that becomes standard in the last decades is \textit{Dirac's (or the invariant) formalism}, which uses an infinite separable Hilbert space, linear operators acting on this space, and Dirac's bra and ket notation \cite{Gieres}.\\

\noindent As it is known, separability of a Hilbert space $\mathcal{H}$ means that $\mathcal{H}$ admits an orthonormal basis consisting of a denumerable (countable) family of vectors. In view of that, let the set of vectors $\{\left|\!\left.{\Phi}_n \!\right.\right\rangle \equiv {\left|\!\left.{n} \!\right.\right\rangle}_{n\in \mathbb{N}}\}$ satisfying the orthonormalization relation\\

\begin{equation} \label{Orthonormalization} 
   \begin{array}{cl}
      \forall n, m \in \mathbb{N}:
      &
      \langle n\! \left|\!\left. m \!\right.\right\rangle = {\delta}_{nm}
      \;\;\;\; 
   \end{array}  
\end{equation}
\smallskip

\noindent and the closure (or completeness) relation\\

\begin{equation} \label{Closure1} 
        \sum^{\infty}_{n=1} 
                                               \left|\!\left. n \!\right.\right\rangle \!\!
                                               \langle n | 
    = 
    \mathbbm{1}_{\infty}
   \;\;\;\;  ,
\end{equation}
\smallskip

\noindent be an orthonormal basis of $\mathcal{H}$, then any vector $\left|\!\left. {\Psi} \!\right.\right\rangle$ of the separable Hilbert space $\mathcal{H}$ can be presented as the expansion\\

\begin{equation} \label{Expansion} 
        \left|\!\left. {\Psi} \!\right.\right\rangle
        =
       \mathbbm{1}_{\infty} \! \left|\!\left. {\Psi} \!\right.\right\rangle
        =
        \sum^{\infty}_{n=1} 
                                               \left|\!\left. n \!\right.\right\rangle \!\!
                                               \langle n\! \left|\!\left. {\Psi} \!\right.\right\rangle 
   \;\;\;\;  ,
\end{equation}
\smallskip

\noindent where the elements ${\Psi}_n \equiv \langle n\! \left|\!\left. {\Psi} \!\right.\right\rangle \in \mathbb{C}$ define the numerical representation of $\left|\!\left. {\Psi} \!\right.\right\rangle$.\\

\noindent A linear operator $L$ on the Hilbert space $\mathcal{H}$ is a linear map\\

\begin{equation} \label{Operator} 
     \begin{array}{l l}
         L: &  \mathcal{D}(L) \rightarrow \mathcal{H}                                                                                     \;\;\;\;\;   , \\
             &   \left|\!\left. {\Psi} \!\right.\right\rangle  \rightarrow   L \!\left|\!\left. {\Psi} \!\right.\right\rangle       \;\;\;\;   .
      \end{array}
      \;\;\;\;   
\end{equation}
\smallskip

\noindent where $\mathcal{D}(L)$ denotes the domain of definition of this operator.\footnote{If the spectrum of $L$ is not bounded, then the domain of definition of $L$ cannot be all of $\mathcal{H}$, that is, $\mathcal{D}(L) \neq \mathcal{H}$. Furthermore, if the spectrum of the operator $L$ contains a continuous part, then the corresponding eigenvectors do not belong to $\mathcal{H}$, but to a larger space.\vspace{5pt}} Using the closure relation (\ref{Closure1}), any linear operator $L$ acting on the infinite separable Hilbert space $\mathcal{H}$ can be written down as\\

\begin{equation} \label{Matrix} 
        L
        =
       \mathbbm{1}_{\infty}  L  \mathbbm{1}_{\infty}
        =
        \sum^{\infty}_{n,m} \langle n\! \left|\!\left. L |m \!\right.\right\rangle
                                               \left|\!\left. n \!\right.\right\rangle \!\!
                                               \langle m | 
   \;\;\;\;  ,
\end{equation}
\smallskip

\noindent where the elements $L_{nm}  \equiv \langle n\! \left|\!\left. L |m \!\right.\right\rangle \in \mathbb{C}$ specify the numerical (i.e., matrix) representation of $L$.\\

\noindent The equations (\ref{Expansion}) and (\ref{Matrix}) serve to convert between the infinite Hilbert space $\mathcal{H}$ representation and the number-based (that is, matrix/vector-based) representation. These equations are at the center of what it means to find \textit{a computational representation} of a quantum-mechanical problem. Hence, non-recursivity of the quantum mathematical framework can clearly be grasped from these equations alone.\\

\noindent First thing, one cannot explicitly write down (which means, construct) a numerical vector $\overrightarrow{{\Psi}_{\infty}} \!=\! ({\Psi}_1, \dots, {\Psi}_n, \dots)$ or a numerical matrix ${\mathsf{L}}_{\infty}\!=\!{(L_{nm})}_{nm}^{\infty}$ because these objects contain an infinite number of elements. Furthermore, operations on these numerical objects such as the square norm of a vector, i.e., $\|\overrightarrow{{\Psi}_{\infty}}\|^2 = \sum_{n=1}^{\infty} |{\Psi}_n|^2$, or the matrix trace, i.e., $\mathrm{Tr} ({\mathsf{L}}_{\infty}) = \sum_{n=1}^{\infty}L_{nn}$, may require infinite summations, and therefore in general cannot be defined \cite{Halmos, Shivakumar}, (hence, computed). More importantly, any attempt to truncate those infinite vectors and matrices to a finite order (so as to make them explicit and operations on them well defined) immediately leads to a conflict with \textit{the canonical commutation relation} (CCR) – the fundamental relation between canonical conjugate quantities in quantum mechanics.\\

\noindent To be sure, let us assume that CCR is satisfied by the operators of momentum $P$ and position $Q$, which can be represented by the corresponding finite $N \times N$ matrices ${\mathsf{P}}\!_{N}$ and ${\mathsf{Q}}_{N}$:\\

\begin{equation} \label{CCR} 
       [{\mathsf{P}}\!_{N},{\mathsf{Q}}_{N}] = -i \hbar {\mathsf{1}}_{N} 
   \;\;\;\;  ,
\end{equation}
\smallskip

\noindent where ${\mathsf{1}}_{N}={({\delta}_{nm})}_{nm}^{N}$ is the $N \times N$ identity matrix representing the identity operator $\mathbbm{1}_N$. Taking the trace of the left-hand side of the relation (\ref{CCR}), we immediately find\\

\begin{equation} \label{Tr_left} 
       \mathrm{Tr} [{\mathsf{P}}\!_{N},{\mathsf{Q}}_{N}]
       \equiv
       \mathrm{Tr} \!\left(
                                          {\mathsf{P}}\!_{N}{\mathsf{Q}}_{N}
                                          -
                                         {\mathsf{Q}}_{N}{\mathsf{P}}\!_{N} 
                              \right)
        = 0
   \;\;\;\;   
\end{equation}
\smallskip

\noindent since the trace of the product of two square matrices ${\mathsf{P}}\!_{N}$ and ${\mathsf{Q}}_{N}$ is well defined and independent of the order of multiplication. However, the trace of the right hand side of (\ref{CCR}) gives a contradictory result\\

\begin{equation} \label{Tr_right} 
       \mathrm{Tr} \!\left(-i \hbar {\mathsf{1}}_{N} \right) = -i \hbar N
   \;\;\;\;  .
\end{equation}
\smallskip

\noindent As follows, the assumption (\ref{CCR}) is wrong. This means that quantum mechanics can be formulated only on an infinitely dimensional Hilbert space $\mathcal{H}$: On this space, the trace is not a generally well defined operation, in particular, the trace of the identity matrix ${\mathsf{1}}_{\infty}$ representing the identity operator $\mathbbm{1}_{\infty}$, i.e.,  $\mathrm{Tr} \!\left({\mathsf{1}}_{\infty} \right) = \sum_{n=1}^{\infty}{\delta}_{nn}$, does not exist and so it is impossible to deduce a contradiction from CCR $[{\mathsf{P}}\!_{\infty},{\mathsf{Q}}_{\infty}] = -i \hbar {\mathsf{1}}_{\infty}$.\\

\noindent On that account, the mathematical foundations of quantum mechanics can be regarded as \textit{necessarily non-constructive in the sense of constructive recursive mathematics}.\footnote{A general proof (not based on the computational representation of quantum mechanics), which demonstrates that the quantum world is in fact a “hostile environment” for many (if not all) varieties of constructivist mathematics, is given in the paper \cite{Hellman}. As it is argued there, from a thoroughgoing constructivist point of view, unbounded linear Hermitian operators in an infinite Hilbert space cannot be even recognizable as mathematical objects. Also, deserving of mentioning are the investigations \cite{Myrvold94, Myrvold95}, which show that, due to the use of unbounded and thus discontinuous operators in an infinite Hilbert space, noncomputable and hence nonconstructive objects are unavoidable in the quantum theory. Even though it is still unclear whether or not at least some form(s) (albeit modified) of constructivist mathematics can be applied to develop a full counterpart of quantum formalism constructive in a sense, it is crucial to our investigation that \textit{such a counterpart cannot be recursive}.\vspace{5pt}}\\

\section{Constructive substitutes for quantum formalism}\label{Substitutes}

\noindent Nonetheless, the irremovable breakdown of computability in quantum theory cannot preclude a constructive substitute for quantum formalism, at least in some situations.\\

\noindent Let the identity operator $\mathbbm{1}_{\infty}$ acting on the infinitely dimensional Hilbert space $\mathcal{H}$ be formally presented as the sum of two operators $\mathbbm{1}_N$ and  $\mathbbm{X}_{\infty}$, where $\mathbbm{1}_N$ is the projector onto some finite subspace of $\mathcal{H}$, $\mathbb{C}^N={\{\left|\!\left. n \!\right.\right\rangle \}_{n=1}^N} \subset \mathcal{H}$, and $\mathbbm{X}_{\infty}$ is the projector onto the reminder of $\mathcal{H}$, that is, on the relative complement $\mathcal{H} \setminus \mathbb{C}^N$. Then, substituting operator $\mathbbm{1}_{\infty}$ for the sum $\mathbbm{1}_N + \mathbbm{X}_{\infty}$ in the equations (\ref{Expansion}) and (\ref{Matrix}), one gets\\

\begin{equation} \label{Expansion_sub} 
        \left|\!\left. {\Psi} \!\right.\right\rangle
        =
        \sum^{N}_{n=1}  {\Psi}_n \! 
                                               \left|\!\left. n \!\right.\right\rangle
        +
        \mathbbm{X}_{\infty} \! \left|\!\left. {\Psi} \!\right.\right\rangle          
   \;\;\;\;   
\end{equation}
\smallskip

\noindent and\\

\begin{equation} \label{Matrix_sub} 
        L
        =
        \sum^{N}_{n,m} L_{nm}
                                               \left|\!\left. n \!\right.\right\rangle \!\!
                                               \langle m | 
        +
        \mathbbm{1}_N L \mathbbm{X}_{\infty} 
        +
        \mathbbm{X}_{\infty} L \mathbbm{1}_N
        +
         \mathbbm{X}_{\infty} L \mathbbm{X}_{\infty} 
   \;\;\;\;  .
\end{equation}
\smallskip

\noindent As one can see, constructivism of quantum formalism (to be exact, a constructive substitute for quantum formalism) can be guaranteed in every situation, in which all the terms in the equations (\ref{Expansion_sub}) and (\ref{Matrix_sub}) involving the projector $\mathbbm{X}_{\infty}$ can be safely neglected.\\

\noindent It is reasonable to state that in any experiment on atomic or subatomic particles, which is performed within a region limited along the axes of the parameter space (such as energy, size or the number of the particles involved in the experiment), constructivization of Dirac's formalism – i.e., the replacement of an infinite Hilbert space $\mathcal{H}$ by a finite dimensional vector space $\mathbb{C}^N$ – \textit{can be justifiable in physical terms}.\\

\noindent For example, let us consider a single subatomic particle of mass $m$ moving along a straight line $x$. Suppose that in an experiment carried out on this particle, its otherwise free movement is restricted to a finite region $x \in [0,a]$ (where $a>0$) such that the particle can rarely be detected outside that region. Also, suppose that the considered experiment only involves energies of several million electronvolts or less (such as the arrangement of electrons in an atom or a solid) and therefore high energies of the observed particle can hardly be registered. In such circumstances, we can safely allow only excitation energies $E_n \le E_N$ of the particle that are below some maximal level $E_N$\\

\begin{equation} \label{Maximal_level} 
        E_N
        =
        \frac{N^2 {\pi}^2 {\hbar}^2} {2ma^2} 
   \;\;\;\;   
\end{equation}
\smallskip

\noindent and thereby can restrict the orthonormal basis of the infinite Hilbert space ${\{\left|\!\left. n \!\right.\right\rangle \}_{n=1}^{\infty}}$ to the truncated, or \textit{computational}, basis ${\{\left|\!\left. n \!\right.\right\rangle \}_{n=1}^N}$.\\

\noindent In this basis, all the expressions of Dirac's formalism become well defined since any operator $L$ on $\mathbb{C}^N$ (which now can be expressed in terms of a finite $N \times N$ matrix) and its adjoint (i.e., the Hermitian conjugate matrix) are defined on the entire (truncated) Hilbert space $\mathcal{H} = \mathbb{C}^N$, i.e., $\mathcal{D}(L)=\mathbb{C}^N$. On the other hand, only those mathematical expressions may possibly be effectively calculable that are well defined. Hence, in the truncated (computational) basis, the expressions of quantum formalism may well be constructive.\\

\noindent For instance, one would need only a finite number of steps in order to construct (to any given precision of the values of $\pi$, $\hbar$, $m$ and $a$) the numerical matrix ${\mathsf{H}}_{N}\!=\!{(H_{nm})}_{nm}^{N}$ with elements\\

\begin{equation} \label{Hamiltonian} 
        H_{nm}
        =
        \langle n\! \left|\!\left. H |m \!\right.\right\rangle
        =
        \frac{n^2 {\pi}^2 {\hbar}^2} {2ma^2} \times {\delta}_{nm}
   \;\;\;\;   
\end{equation}
\smallskip

\noindent that (approximately) describes the Hamiltonian operator $H$ of the particle under consideration or to (again, approximately) exhibit an arbitrary wavefunction $\Psi (x)= \langle x\! \left|\!\left. {\Psi} \!\right.\right\rangle$ of this particle at any given point $x$ of the region $[0,a]$\\

\begin{equation} \label{Wavefunction} 
        \Psi (x)
        \approx
        \sum^{N}_{n=1} 
                                               \langle x\! \left|\!\left.      n    \!\right.\right\rangle \! 
                                               \langle n\! \left|\!\left. {\Psi} \!\right.\right\rangle
         =
         \sqrt{\frac {2}{a}} \sum^{N}_{n=1} c_n \sin \left( {\frac {n \pi x}{a}} \right)
   \;\;\;\;  ,
\end{equation}
\smallskip

\noindent where $\left|c_n\right|^2$ is the probability of measuring the particle's energy to be $E_n$.\\

\noindent In another example, let us consider a system of $N$ microscopic (e.g., atomic or subatomic) particles moving at less than a relativistic velocity in a confined area ${\Omega}_3=\left({\Omega}_x,{\Omega}_y,{\Omega}_z\right)$ of three-dimensional space. To characterize such a system would require a wave function  $\Psi_{3N}(Q)= {\langle Q\! \left|\!\left. {\Psi} \!\right.\right\rangle}_{3N}$, where $Q$ is the vector representing $3N$ spatial coordinates of the system\\

\begin{equation} \label{Configuration} 
        Q
        \equiv
        \left(
        x_1, y_1, z_1, \dots, x_N, y_N, z_N
        \right)
   \;\;\;\;   
\end{equation}
\smallskip

\noindent and the state (configuration) $\left|\!\left. Q \!\right.\right\rangle \equiv \left|\!\left. {\Psi}_Q \!\right.\right\rangle$ (i.e., the spatial state of the state vector $\left|\!\left. {\Psi} \!\right.\right\rangle$ associated with the position state $\left|\!\left. Q_0 \!\right.\right\rangle \equiv \left|\!\left. {\Psi}_{Q_0} \!\right.\right\rangle$) is the tensor product $\left|\!\left. Q \!\right.\right\rangle =\left|\!\left. {x_1} \!\right.\right\rangle\! \otimes \!\left|\!\left. {y_1} \!\right.\right\rangle\! \otimes \!\left|\!\left. {z_1} \!\right.\right\rangle\! \otimes \!\dots \otimes \!\left|\!\left. {z_N} \!\right.\right\rangle$. Let the Hamiltonian of this system take the form that does not depend on the time $t$\\

\begin{equation} \label{Hamiltonian_N} 
        H
        = 
        \sum^{N}_{n=1}
        \frac{{\hbar}^2} {2m_n}
        \left(
                          \frac{{\partial}^2}{\partial{x^2_n}} + \frac{{\partial}^2}{\partial{y^2_n}} +\frac{{\partial}^2}{\partial{z^2_n}}
        \right)
        + V(Q)
   \;\;\;\;  ,
\end{equation}
\smallskip

\noindent where $m_n$ is the mass of the $n^{\mathrm{th}}$  particle and $V(Q)$ is the system's potential energy. Then, the matrix element of the propagator $U(t_0,t)=\exp{(-iH(t-t_0)/\hbar)}$ representing the formal solution to the time-dependent Schrödinger equation of this system would be defined by the expression\\

\begin{equation} \label{Amplitude} 
        A
        = 
        \left \langle \! Q^{\prime} \middle| \left. {\mathrm{e}^{\frac{-iH}{\hbar}(t-t_0)}}  \middle|Q^{{\prime}{\prime}} \! \right.\right\rangle
   \;\;\;\;  ,
\end{equation}
\smallskip

\noindent that gives the amplitude $A$ with which the configuration $Q'$  turns into the configuration $Q''$ during the evolution period $(t-t_0)$.\\

\noindent Let us regard the configurations $Q^{\prime}$ and $Q^{{\prime}{\prime}}$ as the starting and ending points of a path running through the configuration space of the system. Then, dividing the evolution period $t-t_0$ into a (large) number $K \in \mathbb{N}$ of intervals and presenting the starting and ending points of this path as $Q(t_0)=Q^{\prime}$ and $Q(t)=Q^{{\prime}{\prime}}$, we find that in the case where all the particles have the same mass (i.e., $\forall n \in \mathbb{N}: m_n \equiv m)$ the amplitude (\ref{Amplitude}) is given by the formula\\

\begin{equation} \label{Amplitude2} 
        A
        \approx
        \left( - \frac{imK}{2\pi \hbar t} \right)^{\frac{3NK}{2}}
        \int_{{\Omega}_{3N} \in \mathbb{R}^{3N}} 
              \mathrm{e}^{\frac{iS}{\hbar}} 
              \mathrm{d}^{3N}Q_1 \dots \mathrm{d}^{3N}Q_k \dots \mathrm{d}^{3N}Q_{K-1}
   \;\;\;\;  ,
\end{equation}
\smallskip

\noindent in which every $\mathrm{d}^{3N}Q_k$ is the $3N$-dimensional volume differential of the $3N$-tuple  $Q_k$ defined by\\

\begin{equation} \label{Q_k} 
        \forall k \in [0,K]: \;\;\;\;
        Q_k \equiv Q({\tau}_k) \equiv Q \left( t_0 + k \frac{t-t_0}{K} \right)
   \;\;\;\;   
\end{equation}
\smallskip

\noindent and the term $S$ stands for the action of the path (the sequence) $Q_0,…,Q_k,…,Q_K$\\

\begin{equation} \label{Action} 
        S = \frac{m}{2}\frac{K}{t}
        \sum^K_{k=1} \|Q_k-Q_{k-1}\|^2
        -
        \frac{t}{K} \left(
                                 \frac{1}{2}V(Q_0)+\sum^{K-1}_{k=1}V(Q_k)+\frac{1}{2}V(Q_K)
                         \right)
   \;\;\;\;  .
\end{equation}
\smallskip

\noindent Formally, the expression (\ref{Amplitude2}) can be written as an integral over all (continuous and differentiable) paths $Q(\tau_k)$ running through $3N$-dimensional configuration space of the system from the configuration $Q^{\prime}$ to the configuration $Q^{{\prime}{\prime}}$:\\

\begin{equation} \label{Path_integral} 
        A
        =
        \int_{Q^{\prime}}^{Q^{{\prime}{\prime}}} \!\!\mathrm{e}^{\frac{iS[Q(\cdot)]}{\hbar}} \mathcal{D}_{t-t_0}[Q(\cdot)]
   \;\;\;\;   
\end{equation}
\smallskip

\noindent (again, the last expression should be viewed as a notation for the more precise expression (\ref{Amplitude2}) as the number of the intervals $K$ tends to infinity; see technical details in \cite{MacKenzie} and \cite{Schmied}.)\\

\noindent If we assume that the number of the constituent particles $N$ of the given system is upper bounded, i.e., $N<\infty$ (which is a natural assumption for most physical systems), then the identity operator $\mathbbm{1}_{3N}$, which integrates over all coordinates of $N$ particles\\

\begin{equation} \label{Identity_3N} 
        \mathbbm{1}_{3N}
        \equiv
        \int_{{\Omega}_{3N} \in \mathbb{R}^{3N}}
              \!\left|\left. Q \!\right.\right\rangle \!\! \langle Q | \mathrm{d}^{3N}Q
        =
        \int_{{\Omega}_{x_1} \in \mathbb{R}}
              \!\left|\left. x_1 \!\right.\right\rangle \!\! \langle x_1 | \mathrm{d}x_1
        \otimes \dots \otimes
        \int_{{\Omega}_{z_N} \in \mathbb{R}} 
              \!\left|\left. z_N \!\right.\right\rangle \!\! \langle z_N | \mathrm{d}z_N
   \;\;\;\;  ,
\end{equation}
\smallskip

\noindent will become potentially realizable. This means that we may regard (\ref{Identity_3N}) as a well-defined operation since for each $q_n=(x_n,y_n,z_n)$, where $n \in [1,N]$, the operation $\mathbbm{1} \equiv \int_{{\Omega}_{q_n} \in \mathbb{R}}\!\left|\left. q_n \!\right.\right\rangle \!\! \langle q_n | \mathrm{d}q_n$ is well-defined and in such manner $\mathbbm{1}_{3N}$ can be completed for an arbitrary large number $N$. Accordingly, the path integration in (\ref{Path_integral}) can be calculated effectively to any desired precision, i.e., for any number of the intervals $K$ (though, this path integration might be unfeasible for any reasonably sized problem).\\

\section{A constructivist approach to the universal wave function}\label{Universal_wavefunction}

\noindent As it is known, there is no commonly accepted answer to the question what a wave function of the universe means.\footnote{Suffice it to say that it is even unclear as to how to account for the probabilistic content of the wave function of the universe; see, for example, the discussion in \cite{Hemmo}.\vspace{5pt}} Probably, it is so because such a question touches upon deep unresolved uncertainties in the interpretation of quantum mechanics. But no matter what the controversy over the meaning of this function is, one thing is in no doubt: It is generally assumed as true that the wave function of the universe must contain all of the information about the geometry and matter content of the universe in such a way that all of physics follows from this function alone. It is also certain that the environment is presumed to be a part of the universe, and therefore one can no longer make the conventional split between a ‘system’ and its ‘environment’ regarding the universe. Together these two things imply that in the case of the quantum model of the entire universe constructivization of Dirac's formalism \textit{cannot be justified}.\\

\noindent Let us elaborate upon this statement. As the environment is part of the universe, the limitations along the axes of the parameter space of the universe (such as the energy of the universe, the size of the universe or the amount of the microscopic particles in the universe) would suggest that there exist fundamental physical constants that signify the maximal limits for the corresponding parameters (e.g., the maximum energy limit of the universe, the maximal size of the universe or the maximal amount of the microscopic particles in the universe). However, those fundamental limits are nonexistent in the known physical laws. Accordingly, when applying quantum formalism to the totality of existence, we cannot rationalize the replacement of an infinite Hilbert space by a finitely dimensional vector space.\\

\noindent For example, it is possible that the universe – while continuing to expand – is already \textit{spatially infinite}.\footnote{According to the commonly accepted paradigm, the observed universe evolved in a finite time from a dense singular state, before which classical space and time did not exist. However, as it is argued in the papers \cite{Aguirre, Guth}, self-consistent, geodesically complete, and physically sensible steady-state (SS) eternally inflating universe, based on the flat slicing of de Sitter space, is also possible. In the SS, the universe always has and always will exist in a state statistically like its current one, and time has no beginning. Needless to say, the SS cosmology is appealing because it avoids an initial singularity, has no beginning of time, and does not require an initial condition for the universe.\vspace{5pt}} Provided that the volume of a universe grows in proportion with the total of the atomic and subatomic particles contained within it (under the conditions of approximate homogeneity and isotropy of the matter distribution on the largest scales), we can determine that in the case of a spatially infinite universe the maximum limit of the number of the microscopic particles $N$ does not exist, in other words, the amount $N$ is unbounded and forms an actual infinity.\\

\noindent For a finite amount $N$, the inner product of the vector $\left|\Psi \right\rangle\!_{3N}$ with itself can be obtained using the continuous closure relation (\ref{Identity_3N}):\\

\begin{equation} \label{Inner_product} 
        _{3N}\!\!\left \langle \Psi \middle| \Psi \! \right\rangle\!_{3N}
        =
        _{3N}\!\! \left \langle \Psi \middle| \left. \mathbbm{1}_{3N} \middle|\Psi \! \right.\right\rangle\!_{3N}
        =
        \int_{\mathbb{R}^{3N}} \! \left|\Psi_{3N}(Q)\right|^2 \mathrm{d}^{3N}Q
   \;\;\;\;  .
\end{equation}
\smallskip

\noindent Born’s probabilistic interpretation of the wave function requires that this product must be unity; thus, the wave function $\Psi_{3N}(Q)$ must be square integrable with respect to the spatial coordinates $Q$, that is, $\Psi_{3N}(Q) \in L^2(\mathbb{R}^{3N},\mathrm{d}^{3N}Q)$, where\\

\begin{equation} \label{L2_space} 
        L^2\left(\mathbb{R}^{3N},\mathrm{d}^{3N}Q \right)
        =
        \left\{
        f: \mathbb{R}^{3N} \to \mathbb{C} \;
        \middle|
        \; \int_{\mathbb{R}^{3N}} \! \left|f(Q)\right|^2 \mathrm{d}^{3N}Q
        \right\}
   \;\;\;\;   
\end{equation}
\smallskip

\noindent is the space of square integrable functions.\\

\noindent It is reasonable to assume that for most macroscopic systems (i.e., ones that contain a large number of microscopic particles $N \ggg 1$), adding or removing a few particles to or from a system will result in rather small changes in the state of the system and, thus, in the state vector of the system $\left|\Psi \right\rangle_{3N}$. Consequently, the state vector $\left|\Psi \right\rangle_{3N}$ can be considered continuous along the parameter $N$ and, subsequently, the infinitely dimensional state vector $\left|\Psi \right\rangle_{\infty}$ (describing an infinite universe) can be deemed to be the limit of the $3N$ dimensional state vector $\left|\Psi \right\rangle_{3N}$ as $N$ tends to infinity,\footnote{Putting it differently, because an essential feature of macroscopic assemblies of microscopic particles is that the state equations are size independent, one can naturally arrive at an idealization of an infinite universe as an infinite-volume limit of increasingly large finite systems with constant density.\vspace{5pt}} that is,\\

\begin{equation} \label{Limit} 
        \left|\Psi \right\rangle_{\infty}
        =
        \lim_{N \to \infty} \left|\Psi \right\rangle_{3N}
   \;\;\;\;  .
\end{equation}
\smallskip

\noindent Like so, we can write $\left\langle Q \middle| \Psi \right\rangle_{\infty} = \lim_{N \to \infty} \left\langle Q \middle| \Psi \right\rangle_{3N}$, or $\Psi_{\infty}(Q) = \lim_{N \to \infty} \Psi_{3N}(Q)$, and\\

\begin{equation} \label{Inner_product_limit} 
        \int_{\mathbb{R}^{\infty}} \! \left|\Psi_{\infty}(Q)\right|^2 \mathrm{d}^{\infty}Q
        =
        \lim_{N \to \infty}
        \int_{\mathbb{R}^{3N}} \! \left|\Psi_{3N}(Q)\right|^2 \mathrm{d}^{3N}Q
   \;\;\;\;  .
\end{equation}
\smallskip

\noindent Let us consider the case where $\left|\Psi_{3N}(Q)\right|^2$ takes the form of a Gaussian function\\

\begin{equation} \label{Gaussian} 
        \left|\Psi_{3N}(Q,a)\right|^2
        =
        \exp(-aQ^2)
   \;\;\;\;  ,
\end{equation}
\smallskip

\noindent in which $a$ is an arbitrary positive parameter and $Q^2=\sum_{n=1}^N \left(x_n^2+y_n^2+z_n^2 \right)$. Substituting (\ref{Gaussian}) into (\ref{Inner_product}), we find that the inner product of the wave function $\Psi_{3N}(Q,a)=\left\langle Q \middle| \Psi (Q,a) \right\rangle_{3N}$ with itself is a function continuous along the parameter $a$\\

\begin{equation} \label{Inner_product_3N} 
        f(a)
        =
        \int_{\mathbb{R}^{3N}} \! \left|\Psi_{3N}(Q,a)\right|^2 \mathrm{d}^{3N}Q
        =
        \left(\frac{\pi}{a}\right)^{\frac{3N}{2}}
   \;\;\;\;  .
\end{equation}
\smallskip

\noindent This means that by taking the parameter $a$ to be \textit{sufficiently close} to $\pi$, the integral in (\ref{Inner_product_3N}) can be made as close as desired to unity.\\

\noindent In contrast to (\ref{Inner_product_3N}), the substitution of (\ref{Gaussian}) into (\ref{Inner_product_limit}) gives a discontinuous at $a=\pi$ function\\

\begin{equation} \label{Inner_product_inf} 
        f(a)
        =
        \int_{\mathbb{R}^{\infty}} \! \left|\Psi_{\infty}(Q,a)\right|^2 \mathrm{d}^{\infty}Q
        =
       \left\{
              \!\!{
              \begin{array}{l l}
              0,       & a>\pi \\
              1,       & a=\pi \\
              \infty, & a \in (0,\pi) \\
              \end{array}
              }
       \right.
   \;\;\;\;  .
\end{equation}
\smallskip

\noindent which means that the non-zero wave function $\Psi_{\infty}(Q,a)=\left\langle Q \middle| \Psi (Q,a) \right\rangle_{\infty}$ will belong to the space of square integrable functions $L^2\left(\mathbb{R}^{\infty},\mathrm{d}^{\infty}Q \right)$ only if the parameter $a$ is \textit{exactly equal} to the number $\pi$.\\

\noindent On the other hand, being an irrational number, $\pi$ has an infinite number of digits in its decimal representation. Hence, to get the parameter $a$ to be exactly equal to $\pi$ would take an infinite number of steps to accomplish because it would require calculating the number $\pi$ to an infinite precision.\footnote{No ratio of positive integers (capable of being calculated in a finite number of steps) can be the exact value of $\pi$; see the proof in \cite{Arndt}.\vspace{5pt}} As follows, the particular case of the statement (\ref{Limit}) concerning the existence of the normalized infinitely dimensional state vector $\left|\Psi (a) \right\rangle_{\infty} = \lim_{N \to \infty} \left|\Psi (a) \right\rangle_{3N}$ is constructively impossible to prove since it would imply an effectively impossible procedure for calculating the exact value of $\pi$.\\

\noindent Thus, taking into consideration this counterexample, the normalized wave function describing an arbitrary universe (or an arbitrary configuration of the universe) cannot be regarded as a constructive entity.\\

\section{Discussion}\label{Discussion}

\noindent Let us briefly examine possible arguments against the last inference.\\

\subsection{The wave function of the universe can be interpreted only in an approximation}\label{Approximation}

\noindent To begin with, one may simply rebuff such an inference saying that nonconstructivism of the quantum model of the entire universe is of no concern for quantum formalism and quantum cosmology since a model of that particular kind can anyway be considered only in some limit (i.e., an approximation) that effectively reduces all the universe's degrees of freedom to just a few.\footnote{A typical example of such a `limited' consideration is the Hartle–Hawking state (i.e., the wave function of the universe) satisfying the Wheeler–DeWitt equation defined in mini-superspaces \cite{Hartle}.\vspace{5pt}}\\

\noindent However, a stance like this leads to a conceptual confusion. According to the paper \cite{Kiefer13}, implications for the meaning of the the wave function of the universe should not be derived from the approximations of this function but only from the function itself. This is so because neglecting all but a small number of the degrees of freedom in the configuration space implies a split between a `system' characterized by the observed degrees of freedom, and its (non-interacting or weakly interacting) `environment' which is set apart by the rest, i.e., neglected (unobserved), degrees of freedom. But then again, since by definition, the universe is a closed system, there cannot be unobserved degrees of freedom relating to the `environment' of the universe. Accordingly, in the case of the quantum model describing the entire universe, a decomposition of an infinite Hilbert space into a finite subspace and its (infinite) reminder has no physical meaning.\\

\subsection{Explicit physical finitism}\label{Finitism}

\noindent It is possible to recover a constructivist character of a wave function that describes the entire universe by postulating nonexistence of an actual infinity in the physical world. Undeniably, if no physical parameter were ever to reach infinity, then the wave function of the universe might be algorithmic, thus, constructive (at least, in principle).\\

\noindent And yet, as it was said before, the explicit physical finitism, which puts explicit limits on possible values of all physical parameters, does not follow from the known at present laws of physics, and, hence, its justification and rationalization might imply a new physics.\footnote{If not, then it would be hard for the proposition of explicit physical finitism to answer the charge of \textit{arbitrariness}: Certainly, no matter where the limit for the particular parameter would be drawn, it would be always ad hoc and so perpetually subject to shifting.\vspace{5pt}}\\

\noindent For example, the universe's spatial finitism (and, as a consequence, the existence of the upper bound for the number of the microscopic particles in the universe) may imply a breakdown of relativistic gravity (RG) at the limit of extremely large length scales (which would make eternal expansion of the space impossible). But clearly, one should have some specific a priori reason for believing that RG may break down when pushed far enough along the length axis of the parameter space. And since such an a priori reason does not exist, the assumption of the explicit spatial finitism does not seem realistic (at least, at the moment).\footnote{It is interesting to note that the existence of the upper bound on \textit{the complexity} of the universal wave function could also result in constructiveness of this function. Indeed, according to \cite{Brown}, the complexity of a quantum state ${\left|\!\left. {\Psi} \!\right.\right\rangle}\!_N$ on $N$ qubits is defined as the minimum number of gates (i.e., elementary unitaries) necessary to produce ${\left|\!\left. {\Psi} \!\right.\right\rangle}\!_N$ from a simple reference state $\left|\!\left. {\Psi}\!_0 \!\right.\right\rangle$. The complexity of ${\left|\!\left. {\Psi} \!\right.\right\rangle}\!_N$ in principle depends on all the details of the construction, but more importantly, it is proportional to the number of active degrees of freedom of the system, that is, the number of qubits $N$. So, if the complexity of the state vector describing the universe ${\left|\!\left. {\Psi} \!\right.\right\rangle}\!_N$ were to be limited it would imply the limitation on $N$ and thus the computability of this vector. The problem is that it is unclear why the complexity of the quantum state of the universe should be bounded (and not grow to infinity).\vspace{5pt}}\\

\subsection{Hypercomputing}\label{Hypercomputing}

\noindent Hypothetically, all operations on infinite vectors and matrices (including the square norm of a vector and the trace of a matrix) might be well defined if it were possible to complete the result of infinite summations. Therefore, admitting models of computation that have such capabilities (which are beyond those of a Turing machine – classical and quantum alike \cite{Costa}), the mathematical formalism of quantum mechanics might be allowed to be constructive (in the sense of constructive recursive mathematic) across the whole (unbounded) parameter space.\\

\noindent Aforementioned models of computation are known as \textit{hypercomputers} or \textit{super-Turing machines}. For example, one of them, an ``\textit{accelerated Turing machine}'' \cite{Stannett}, can find the result of infinite summations in a finite amount of time by performing each summation in infinitely short times. Another model, \textit{a BSS machine} \cite{Blum, Bournez}, allows discontinuous functions such as (\ref{Inner_product_inf}), it operates on infinite-precision real numbers and can get the ratio $\pi/a$ to be equal to unity in a single step. Thus, by using these machines, the wave function of the universe can be made to be constructive.\\

\noindent Unfortunately, these machines (together with all other proposed so far models of hypercomputation) do not seem to be physically constructible and reliable.\footnote{For details see the critical analysis \cite{Davis06} that  examines the possibility of computation-like processes transcending the limits imposed by the Church–Turing thesis.\vspace{5pt}} Moreover, in the words of Davis \cite{Davis04}, any ``usable physical representation of an uncomputable function would require a revolutionary new physical theory'', and so if at least some of hypercomputational models were to turn to be real, it would imply evidence that the universe is not as we think of it. For example, a physical realization of the accelerated Turing machine would mean that there exist time intervals shorter than the level of the Planck time widely considered as the scale at which current physical theories fail.\\

\section{Conclusion}\label{Conclusion}

\noindent To conclude, let us discuss how the acknowledgement of non-constructivity of the universal wave function may affect possible interpretations of this function.\\

\noindent As stated in \textit{the proof-conditional criterion of meaningfulness} of radical constructivist programs, a mathematical concept can be meaningfully applicable only if an idealized mathematician (i.e., a constructing intelligence) has a constructive method that shows that it applies. And since there is no constructive method to exhibit the universal wave function in general (i.e., with respect to all possible universes or all possible configurations of the universe), we can conclude that such a function cannot exist (at least, from the mathematical point of view).\\

\noindent One can argue that mathematical constructivism has nothing to do with the foundational problems of quantum mechanics. Above all, the principles of radical constructivist philosophy are inapplicable to the understanding of the universal wave function and the many-world interpretation of quantum mechanics (MWI), which asserts the objective reality of this function \cite{Wallace}.\\

\noindent Definitely, physics cannot limit itself only to the properties that have cognitive significance for finite man. From the viewpoint of radical constructivism, physical science, being a human enterprise, is amenable to human limitations and, as a result, it cannot transcend – just as individuals cannot – the domain of experience. So, to claim that only those mathematical statements in quantum theory can be considered true, for which algorithmic, i.e., computable, proofs are produced, and that a wave function can only exist if an effective method for calculating this function exists, is to confuse the physical reality with a human way of the perception, computation or simulation of that reality. The physical reality does not have any need of being recursive (constructive) and neither does the wave function of the universe.\footnote{Besides, as expected by most logicians, one can construct an undecidable sentence (i.e., such that neither it itself nor its negation is provable) whose physical meaning seems to be hardly questionable. This happens both in classical and in quantum mechanics (see for example investigations by Pitowski \cite{Pitowsky}, Mundici \cite{Mundici} and Svozil \cite{Svozil}). So, the wave function of the entire universe might be uncomputable and yet full of physical meaning.\vspace{5pt}}\\

\noindent Attractive and logical as this point may seem at the first site, it, nevertheless, is equal, if taken seriously,  to an admission of a pure symbolic character of the wave function of the universe.\\

\noindent Let us assume that a wave function that describes the entire universe does exist as a mathematical entity. Then, it would imply that – no matter how complicated this function is – we (or any other finite constructing intelligence) would be able to deduce from it (say, by running a finite procedure that uses basic mathematical operations on this function) the actual values (or the probability distributions) of all physical variables describing the universe.\\

\noindent But, as it has been just demonstrated, unless the number of the universe's degrees of freedom is fundamentally upper bounded (owing to some unknown physical laws) or hypercomputation is physically realizable, the universal wave function is a non-constructive mathematical entity whose calculation may not be completed. So, even if such a function might exist, operations on it could be undefinable and subsequently the only content (if any) one would be able to deduce from this function would be symbolical or metaphysical.\\

\noindent This casts doubt on a physical meaning of the theory of the universal wavefunction, i.e., the MWI. Indeed, proponents of the MWI often call the square absolute value $\left| \Psi(Q) \right|^2$ of the wave function $\Psi(Q)={\langle Q\! \left|\!\left. {\Psi} \!\right.\right\rangle}$ corresponding to a particular configuration of the universe $\left|\!\left. Q \!\right.\right\rangle$ (or a particular universe with the configuration $\left|\!\left. Q \!\right.\right\rangle$) ``the measure of existence'' or ``degree of reality'' \cite{Saunders, Vaidman}. The question, which then naturally arises in connection with the MWI, is this: What is the meaning of the measure of existence of a particular configuration of the universe (or of a particular universe), if the function $\Psi(Q)$ is uncomputable in principle? Putting it differently, who (or what) will possibly be able to know the measure of existence if to calculate it takes an infinite amount of basic operations and consequently an infinite time?\\

\noindent Obviously, the non-constructivity status of $\Psi(Q)$ makes the theory of the universal wavefunction less physical and more philosophical compared to other interpretations of quantum mechanics. So, rewording Bishop, we may say that within the framework of the MWI quantum formalism has evolved into a fine game that has become its own justification.\\

\bibliographystyle{References}
\bibliography{References}

\end{document}